\documentclass[aps, twocolumn,secnumarabic,nobalancelastpage,nofootinbib]{revtex4} %onecolumn,

\usepackage{graphics}      % standard graphics specifications
\usepackage{graphicx}      % alternative graphics specifications
\usepackage{longtable}     % helps with long table options
\usepackage{url}           % for on-line citations
\usepackage{bm,color}            % special 'bold-math' package
\usepackage{amssymb, amsmath, enumerate, theorem, epsfig,subfigure}

\begin{document}

\title{Realizing the $XY$ Hamiltonian in polariton simulators}

\author{Natalia G. Berloff$^{1,2}$, Kirill Kalinin$^1$, Matteo Silva$^{3}$, Wolfgang Langbein$^{4}$ and Pavlos G. Lagoudakis$^{1,3}$}
\email[correspondence address: ]{pavlos.lagoudakis@soton.ac.uk}
\affiliation{$^1$Skolkovo Institute of Science and Technology Novaya St., 100, Skolkovo 143025, Russian Federation}
\affiliation{$^2$Department of Applied Mathematics and Theoretical Physics, University of Cambridge, Cambridge CB3 0WA, United Kingdom }
\affiliation{$^3$Department of Physics and  Astronomy, University of Southampton, Southampton, SO17 1BJ, United Kingdom}
\affiliation{$^4$School of Physics and Astronomy, Cardiff University, The Parade, Cardiff CF24 3AA, United Kingdom}

\date{\today}

\begin{abstract}{Several platforms are currently being explored for simulating physical systems whose complexity increases faster than polynomially with the number of particles or degrees of freedom in the system. Defects and vacancies in semiconductors or dielectric materials \cite{pla,hanson}, magnetic impurities embedded in solid helium \cite{lemeshko13}, atoms in optical lattices \cite{saffman,simon11},  photons \cite{northup}, trapped ions \cite{kim10,lanyon} and superconducting q-bits \cite{corcoles} are among the candidates for predicting the behaviour of spin glasses, spin-liquids, and classical magnetism among other phenomena with practical technological applications. Here we investigate the potential of polariton graphs as an efficient simulator for finding  the global minimum of the $XY$ Hamiltonian. By imprinting polariton condensate lattices of bespoke geometries we show that we can simulate a large variety of systems undergoing the U(1) symmetry breaking transitions. We realise various magnetic phases, such as ferromagnetic, anti-ferromagnetic, and  frustrated spin configurations on unit cells of various lattices: square, triangular, linear and a disordered graph. Our results provide a route to study unconventional superfluids, spin-liquids, Berezinskii-Kosterlitz-Thouless phase transition, classical magnetism among the many systems that are described by the $XY$ Hamiltonian.}

\end{abstract}

\maketitle
Many properties of strongly correlated spin systems, such as spin liquids and unconventional superfluids  are difficult to study as  strong interactions between $n$ particles become intractable  for $n$ as low as $30$ \cite{sandvik2010}.  Feynman envisioned that a quantum simulator  -- a special-purpose analogue processor --  could be used to solve such problems \cite{feymann}. It is expected that quantum simulators would lead to accurate modelling of the dynamics of chemical reactions, motion of electrons in materials, new chemical compounds  and new materials that could not be obtained with classical computers using advanced numerical algorithms \cite{lloyd}. More generally, quantum simulators can be used to solve hard  optimization problems that are at the heart of almost any multicomponent system: new materials for energy, pharmaceuticals, and photosynthesis, among others \cite{qubit}. Many hard optimisation problems do not necessitate a {\it quantum} simulator as only recently realised through a network of optical parametric oscillators (OPOs) that simulated the Ising Hamiltonian of thousands of spins \cite{yamamoto11, yamamoto14}. The Ising model corresponds to the $n=1$ case of the $n$-vector model of classical unit  vector spins ${\bf s}_i$ with the Hamiltonian ${\cal H}_I=-\sum_{ij} J_{ij} {\bf s}_i \cdot { \bf s}_j$, where  $J_{ij}$ is the coupling between the sites labelled $i$ and $j$. For $n=2$ the $n$-vector Hamiltonian becomes the $XY$ Hamiltonian ${\cal H}_{XY}=-\sum_{ij} J_{ij} \cos (\theta_i-\theta_j)$, where we have parameterized unit planar vectors using the polar coordinates ${\bf s}_i=(\cos \theta_i, \sin\theta_i)$. Since ${\cal H}_{XY}$  is invariant under rotation of all spins by the same angle $\theta_i \rightarrow  \theta_i+ \phi$ the XY model is the simplest model that undergoes the $U(1)$ symmetry-breaking transition. As such, it is used to emulate Berezinskii-Kosterlitz-Thouless phase transition and the emergence of a topological order \cite{bkt,bkt2},    topological quantum information processing and storage \cite{nayak08}, and to study  quantum phase transitions, unconventional superfluids, quantum spin models, spin-liquid phases and high-$T_c$  superconductivity. The $XY$ Hamiltonian has been simulated on a triangular lattice of atomic condensates investigating a variety of magnetic phases and frustrated spin configurations \cite{struck11}. Whereas optical lattices offer a scalable platform, they are likely to reach a local rather than global minimum of the Hamiltonian and are limited to sub-$\mu$K temperatures \cite{georgescu14}.

%We envision that polariton graphs offer the scalability of optical lattices, whilst not being limited a sub-$\mu$K temperatures and  are likely to reach the global rather than a local minimum of the XY Hamiltonian \cite{georgescu14}.}
%Whereas atoms in optical lattices have the advantage of scalability \nb{ in comparison with other simulator platforms,} they are likely to reach a local rather than the global minimum of the XY Hamiltonian \cite{georgescu14}.

In this Article, we propose and experimentally demonstrate the use of polariton graphs as a scheme for finding the global minimum of the $XY$ Hamiltonian. Polaritons are the mixed light-matter quasi-particles that are formed in the strong exciton-photon coupling regime in semiconductor microcavities \cite{weisbuch}. Under non-resonant optical excitation, rapid relaxation of carriers and bosonic stimulation result in the formation of a non-equilibrium polariton condensate characterized by a single many-body wave-function \cite{Kasprzak}. Polariton condensates can be imprinted into any two-dimensional graph by spatial modulation of the pumping source, offering the scalability matched only by optical lattices \cite{georgescu14}. Optically injected polariton condensates can potentially be imprinted in multi-site configurations with arbitrary polarisation and density profiles offering the possibility to control the separation distance between sites. Such flexibility allows for unprecedented control of the interaction between neighbouring sites. Due to the finite cavity lifetime, polaritons decay in the form of photons (through quasi-mode coupling \cite{amo2010,ciutiPRB2000}) that carry all information of the corresponding polariton state (energy, momentum, spin and phase). The continuous coupling of polaritons to free photons allows for the in-situ characterisation of static polariton graphs, but more importantly it also allows for the dynamic control of an arbitrary set of sites, whilst measuring in real time the kinetics and phase configuration of the modulated polariton graph.

In a graph of two or more coupled polariton vertices, with increasing excitation density, polariton condensation occurs at the state with the phase-configuration that carries the highest polariton occupation \citep{ohadi14}. This is due to the bosonic character of the condensate formation: the probability of a particle to relax in a particular state grows with the population of that state. Just above condensation threshold a macroscopic coherent state is formed described by the wavefunction $\Psi_g$. $\Psi_g$ can be written as a superposition of the wavefunctions $\Psi_j$ at the sites ${\bf x}_j$ with phase $\theta_j$; that is $\Psi_g\approx\sum_j \Psi_j \exp[i \theta_j]$. Below we will show  that the system of an arbitrary polariton graph condenses into the global minimum of the $XY$ Hamiltonian: ${\cal H}_{XY}=-\sum J_{ij} \cos\theta_{ij}$ where $\theta_{ij}$ is the phase difference between two sites, $\theta_{ij}=\theta_i-\theta_j$ and $J_{ij}$ is the corresponding coupling strength; the latter depends on the density of the sites $i$ and $j$, the distance between them, $d_{ij}=|{\bf x}_i-{\bf x}_j|$, and the outflow condensate wavenumber $k_{c}$, which under non-resonant optical excitation depends on the pumping intensity and profile. The bottom-up approach for the search of the global minimum of the $XY$ Hamiltonian is achievable within the linewidth of the corresponding state similarly to a network of time-multiplexed OPOs \citep{yamamoto14} that guarantees a phase-transition to the global minimum of the Ising Hamiltonian. This is an advantage over classical or quantum annealing techniques, where the global ground state is reached through transitions over metastable excited states (local minima), with an increase of the cost of the search with the size of the system.

\begin{figure}[t]
\centering
  \includegraphics[width=8.6cm]{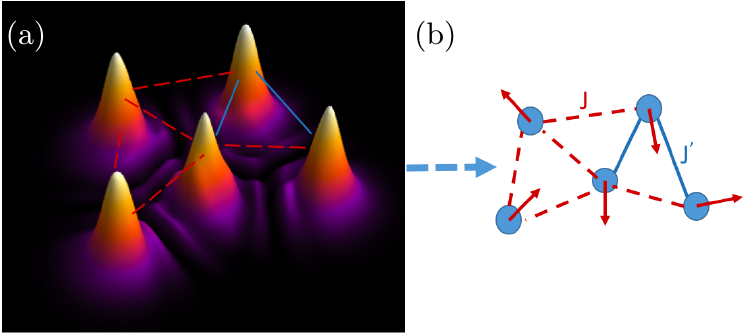}
       \caption{  (a) Schematic of the condensate density map for a five-vertex polariton graph. The sign of the coupling is annotated for some of the edges of the graph: depending on the separation distance between the sites and the outflow wavevector $k_c$ the interactions are either ferromagnetic (solid-blue lines) or anti-ferromagnetic (dashed-red lines). At each vertex ${\bf x}_i$ of the graph polaritons have a local phase $\theta_{i}$ that is mapped to a classical vector spin ${\bf s}_i=(\cos\theta_i,\sin\theta_i)$. (b) the  vertices (blue solid-circles) and edges of the polariton density map depicted in (a), showing the sign of the coupling and the spin vector ${\bf s}_i$ of each vertex.}
     \label{schema}
      \end{figure}

{\it Modelling the phase coupling:} we model the phase coupling in polariton graphs using the complex Ginzburg-Landau equation (cGLE) with a saturable nonlinearity  and energy relaxation \citep{Wouters, Berloff}:
\begin{eqnarray}	
	i \hbar  \frac{\partial \psi}{\partial t} &=& - \frac{\hbar^2}{2m}  \left(1 - i \eta_d {\cal R} \right) \nabla^2\psi + U_0 |\psi|^2 \psi+
	\hbar g_R {\cal R} \psi  \nonumber \\
	  &+&\frac{i\hbar}{2} \biggl(R_R {\cal R} - \gamma_C \biggr) \psi, \label{Initial GL equation}\\
	  \frac{\partial \cal R}{\partial t} &=&  - \left( \gamma_R + R_R |\psi|^2 \right) {\cal R} + P({\bf r}) ,
	\label{Initial Reservoir equation}
\end{eqnarray}
where $\psi$ is the condensate wavefunction, ${\cal R}$ is the density profile of the hot exciton reservoir, $m$ is the polariton effective mass, $U_0$ and  $g_R$ are the strengths of  effective polariton-polariton interaction  and the blue-shift due to interactions with non-condensed particles, respectively, $R_R$ is the rate at which the exciton reservoir feeds the condensate, $\gamma_C$ is the decay rate of condensed polaritons, $\gamma_R$ is the  rate of redistribution of  reservoir excitons between the different energy levels, $\eta_d$ is the energy relaxation coefficient specifying the rate at which gain decreases with increasing energy, and $P$ is the pumping  into the exciton reservoir.  In Eq.~(\ref{Initial Reservoir equation}) we neglected the diffusion of the reservoir  as well as density-density repulsion with the condensate in the view of the large mass of the hot exciton as compared to the mass of the polariton (five orders of magnitude). We non-dimensionalize these equations using
$
        \psi  \rightarrow  \sqrt{\hbar^2  / 2m U_0 l_0^2} \psi,
        {\bf r}  \rightarrow  l_0 {\bf r}, t \rightarrow 2m t l_0^2/ \hbar$ and introducing the notations
  $g = 2 g_R/R_R,$ $\gamma = m \gamma_C l_0^2/ \hbar $,  $ p=m l_0^2 R_R P({\bf r})/ \hbar \gamma_R,\eta = \eta_d \hbar  / mR_R l_0^2,$ and
$ b = R_R \hbar^2  / 2m  l_0^2\gamma_R U_0. $ We choose $l_0=1\mu m$ and
consider the stationary states.

By using the  Madelung transformation $\Psi=\sqrt{\rho}\exp[i S]$ in the dimensionless Eqs. (\ref{Initial GL equation},\ref{Initial Reservoir equation}), where   $\rho=|\psi|^2$,  ${\bf u}=\nabla S$ is the velocity, $S$ is the phase and separating the real and imaginary parts we obtain the mass continuity and the integrated form of the Bernoulli equation which we write for a steady state, and, therefore, introduce the chemical potential $\mu$
\begin{eqnarray}
\mu = - \frac{ \nabla^2\sqrt{\rho}}{\sqrt{\rho}} + {\bf u}^2 + \rho &+& \frac{p({\bf r})}{1 + b \rho} \biggl( g - \eta \frac{\nabla \cdot (\rho {\bf u})}{\rho} \biggr)  ,\label{systemM1} \\
\frac{\nabla\cdot ( \rho {\bf u})}{\rho} =  \frac{p({\bf r})}{1+b\rho} \biggl( 1 &+& \eta \left( \frac{\nabla^2 \sqrt{\rho}}{\sqrt{\rho}} - {\bf u}^2 \right) \biggr) - \gamma.
\label{systemM2}
\end{eqnarray}
First, we consider a single pumping spot with a radially symmetric pumping profile. Asymptotics at large distances from the center of the pump gives the velocity $|u|=k_c=const$  and $\rho\sim \exp[-\gamma r/k_c]r^{-1}.$ From Eq.~(\ref{systemM1}) at infinity, therefore, we obtain
	$\mu=k_c^2 - \gamma^2 / 4 k_c^2$. We can  estimate the chemical potential for a wide pumping spot so that the quantum pressure term $\nabla^2 \sqrt{\rho}/\sqrt{\rho}$ and $u_r$ are insignificant at the pumping center. Under this assumption $\rho_{\max} \approx (p_{\max}-1)/b$ and $\mu \approx(p_{\max}-1)/b + g.$
In \cite{ohadi14} we established experimentally  for the pulsed excitation that the coupling between two pumping spots (a``polariton dyad") can be either in-phase or with a $\pi$ phase difference depending on the outflow wavenumber $k_c$ and the distance between the spots.  Below, in the steady state excitation regime, we obtain a general criterion for the switching between the relative phases. We start by considering the wavefunction of the condensate as the sum of the wavefunctions of individual condensates located at $\pm {\bf d}/2$, where $\pm{\bf d}=(\pm d_{ij},0)$ with the phase difference $\theta_{ij}$:
$
 \tilde{\Psi}(\mathbf{r})\approx
 {\Psi}\left(\mathbf{r}+\frac{\mathbf{d}}{2}\right)
 +e^{i\theta_{ij}}{\Psi}\left(\mathbf{r}-\frac{\mathbf{d}}{2}\right).
$
The  total number of condensed polaritons  can be found in Fourier space as
\begin{equation}
\label{eq:lifetime-integral}
 N  = \int\frac{d{\bf k}}{2\pi^2}
  |\widehat{\Psi}(k)|^{2}[1 +\cos({\bf k}\cdot {\bf d}-\theta_{ij})],
\end{equation}
where $\widehat{\Psi}(k)=2\pi\int_0^\infty\sqrt{\rho(r)} \exp[i k_c r]J_0(kr)r\,dr $ is the Hankel transform of the wavefunction of an individual condensate. We conclude that
\begin{eqnarray}
 N  &=& N_i + N_j + J_{ij}\cos\theta_{ij},\label{N1}\\
  J_{ij}&=&\frac{1}{\pi}\int_0^\infty
  |\widehat{\Psi}(k)|^{2}J_0(kd_{ij})k\, dk.
  \label{J1}
\end{eqnarray}
which in the case of $n$ condensates generalizes to
\begin{equation}
 N  = \sum^n_i N_i + \sum^n_{i<j}J_{ij}\cos\theta_{ij}. \label{nnn}
\end{equation}
The oscillating behaviour of the Bessel function, $J_0(k d_{ij})$, brings about the sign change in the coupling constants $J_{ij}$ depending on  the distance $d_{ij}$.  When $J_{ij}$ is positive the coupling is said to be ferromagnetic and when $J_{ij}$ is negative the coupling is said to be anti-ferromagnetic. The state with the phase configuration that carries the highest number of particles in Eq. (\ref{nnn}) corresponds to the solution that minimises the $XY$ Hamiltonian, ${\cal H}_{XY}=-\sum^n_{i<j}J_{ij}\cos\theta_{ij}$. Between any two polariton nodes the polariton wavefunction forms a standing wave with the density $|\Psi_g|^2\approx\rho_+
 +\rho_-+ 2\sqrt{\rho_+\rho_-}\cos[k_c|x-d_{ij}/2|-k_c |x+d_{ij}/2|-\theta_{ij}]$,  where $x$ is the coordinate along the line that connects the two nodes separated by a distance $d_{ij}$ and $\rho_\pm=\rho(x\pm d_{ij}/2,y)$.  Between two polariton nodes the density oscillates as $1+\cos(2 k_c x + \theta_{ij})$, from which the phase difference $\theta_{ij}$ of a single shot realization can be extracted directly.  In Fig. 1(a) we plot the density of a polariton graph, where for simplicity we have annotated the sign of the coupling for some of the edges of the graph. Depending on the separation distance between the vertices and the outflow wavevector $k_c$ the interactions are either ferromagnetic (solid-blue lines) or anti-ferromagnetic (dashed-red lines). At each vertex ${\bf x}_i$ of the graph polaritons have a local phase $\theta_{i}$, which in the following we map to a classical vector spin ${\bf s}_i=(\cos\theta_i,\sin\theta_i)$. In Fig. 1(b) we show the  vertices of the polariton graph, the edges of Fig.1(a) depicting the sign of the coupling and the spin vector ${\bf s}_i$ of each vertex as calculated from the minimisation of the $XY$ Hamiltonian.

\begin{figure}[t!]
\centering
  \includegraphics[width=8.6cm]{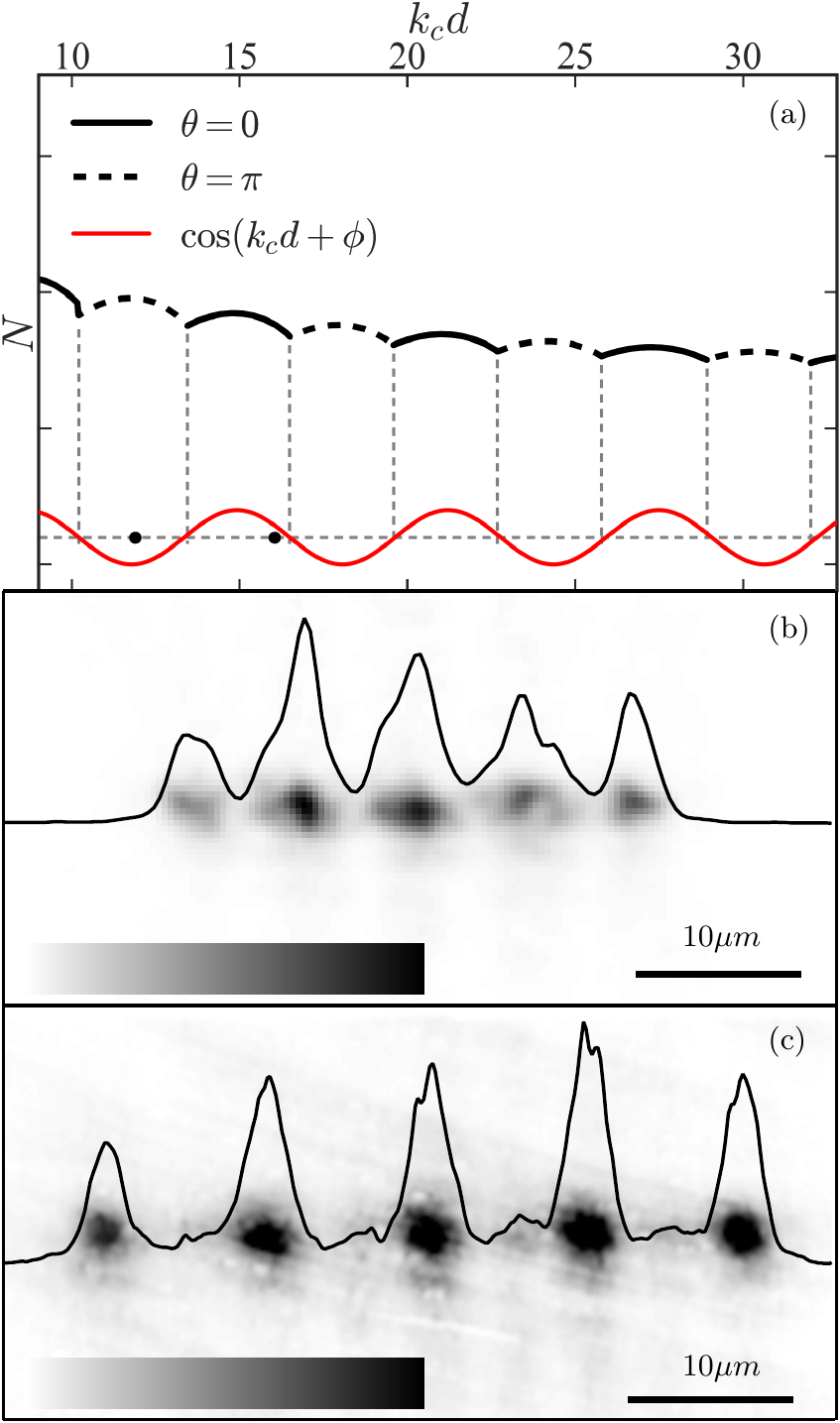}\\ %please fix the SIZE OF FIG2A
%\vspace{-3cm}      	
\caption{ (a) The maximum number of particles, $N$, of a polariton condensate dyad formed under incoherent pumping of two nodes as the function of the product $k_c d$ between the nodes obtained by numerical integration of the cGLE for a fixed $k_c$. The solid black line corresponds to the maximum number of particles in the in-phase ferromagnetic configuration and the dashed black line to the $\pi$-phase difference anti-ferromagnetic configuration. The switching occurs with the periodicity $2\pi/k_c$ as the superimposed graph of $\cos(k_cd + \phi)$ illustrates in red, where $\phi\approx 225^\circ$. (b,c) Experimental realization of an Ising chain of five equidistant polariton nodes with lattice constants of $\sim$ 6.7$\mu m$ and $\sim$9.6$\mu m$ respectively. The  false-grey scale images show the normalised photoluminescence intensity of the real-space tomography at the energy of the condensate; (c) is saturated at 0.5 to increase the visibility of the low intensity fringes between the nodes. The corresponding $k_c d$ (11.9 and 16.04) are shown by two solid circles on (a). }
 \label{NvsD}

\end{figure}

      {\it The Ising polariton chain:} we theoretically describe and experimentally address the minimization of the $XY$ Hamiltonian for the simple case of a linear polariton chain with equal spacing $d=d_{ij}$ between neighbours. For a given $k_c$ with increasing separation distance the coupling between the neighbors, $J_{ij}$, oscillates between negative and positive values.  We approximate the switching of the coupling sign with $\cos(k_c d+\phi)$, where $\phi$ is fixed by the system parameters (see Supp. Mat. for the derivation). 
      %In Ref.\cite{ohadi14}, we investigated the phase coupling mechanism in a polariton dyad, and derived the criterion of coupling in the particular case of expression (\ref{J1}), where each polariton site is sufficiently wide so that $\vert \widehat\Psi(k) \vert^2 \approx  \vert \widehat\Psi(k_c) \vert^2 \delta (k-k_c)$. Under such circumstances, the integral in Eq.(\ref{J1}) evaluates to $J_{ij}= k_c \vert \widehat\Psi(k_c) \vert^2 J_0(k_c d)/\pi$.
      %Thus,  for a given $k_c$ with increasing separation distance the coupling oscillates between negative and positive values. For a finite width we approximate the switching of the coupling sign with $\cos(k_c d+\phi)$, where $\phi$ is fixed by the system parameters (see Supp. Mat. for the derivation).
      %Whereas in Ref. \cite{ohadi14} we used pulsed excitation so as to access the outflow wavevector dependence of $J_{ij}$ for a given separation distance, here, we operate in the steady state regime so as to avoid sign switching in the time domain of the coupling coefficient $J_{ij}$.
      In the steady state excitation regime, we can calculate the maximum particle number of a polariton dyad as a function of the separation distance $d$ by numerically integrating the cGLE to find the solutions of Eqs. (\ref{systemM1}-\ref{systemM2}) for a given pumping profile $p({\bf r})=p_0[\exp(-\alpha |{\bf r-d}/2|^2)+\exp(-\alpha |{\bf r+d}/2|^2)]$ of a characteristic width $\alpha$; the results are shown in Fig. 2(a). The relative phases that realise the maximum particle number switch  periodically between $0$ and $\pi$ with the period $2\pi/k_c$ as shown by superimposing the function $\cos(k_c d + \phi)$ in Fig.2(a); we have used the experimental parameters for the pumping profile and $k_c$ as described in ``Wavevector Tomography'' in Supp. Mat. Where the coupling is ferromagnetic (anti-ferromagnetic) the graph of the maximum number of particles is plotted with a solid (dashed) line. We experimentally address the Ising chain by injecting a linear chain of five equidistant polariton nodes through non-resonant, continuous wave and spatially modulated optical excitation of a multiple InGaAs quantum well semiconductor microcavity that allows for detection of the polariton photoluminescence in the transmission geometry (for the sample description read the ``Microcavity sample" and for the description of the excitation/detection scheme read the ``Experimental setup" in Supp. Mat.). Figures 2(b,c) show the real-space tomography of the photoluminescence intensity at the energy of the condensate from the linear chain with lattice constants of $\sim$ 6.7$\mu m$ and $\sim$9.6$\mu m$ respectively at condensation threshold. The relative phase difference realised between neighbours in the chain is either $\pi$ or zero. The patterns are clearly distinguishable by the number of fringes (density maxima) between the sites: zero or even for  anti-ferromagnetic and odd for ferromagnetic coupling. In Fig.2(a) we have annotated the abscissa with solid circles for each of the two separation distances from which the expected sign of coupling is depicted showing good agreement with the experiment. The observed phase configurations realise the ferromagnetic and anti-ferromagnetic Ising spin chain of the $XY$ model.

%\begin{figure}[]
%\centering
 % \includegraphics[width=3.in]{Figure6,Fourier}
%\caption{ \nb{ The condensate density in Fourier space for the chains of five pumping spots with antiferromagnetic coupling obtained experimentally (a) and numerically (b). }}
% \label{Fourier}
%\end{figure}

\begin{figure}[t]
\centering
  \includegraphics[width=8.6cm]{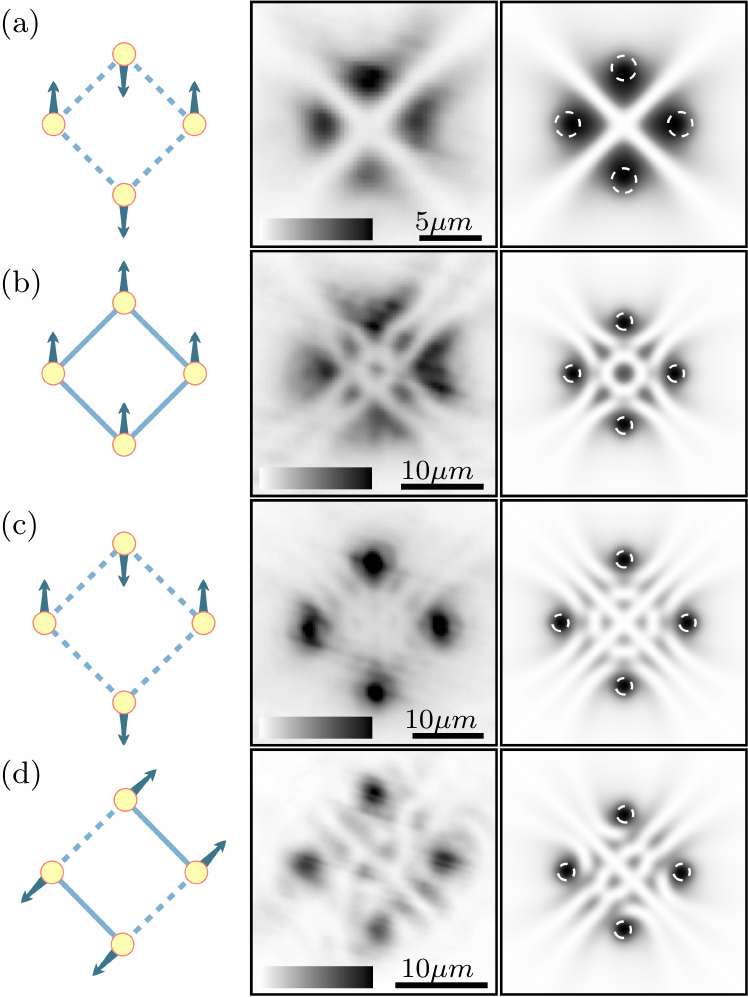}       	
\caption{ Spin configurations of square polariton lattices.  The diagrams of the numerically calculated spins vectors at the pumping sites ${\bf s}_i=(\cos\theta_i,\sin\theta_i)$, the real-space energy tomography of the experimental realisations, and the  averaged condensate densities  of the numerically simulated condensate wavefunctions for several realizations are shown on the left, central and right columns respectively. Solid and dashed blue lines on the spin vector diagrams (left column) indicate ferromagnetic and anti-ferromagnetic coupling, respectively. The  false-grey scale images of the middle column show the normalised photoluminescence intensity of the real-space tomography at the energy of the condensate; (c) is saturated at 0.5 to increase the visibility of the low intensity fringes between the vertices. The configurations shown are some elementary building blocks of square lattices such as  (a,c) anti-ferromagnetic, (b) ferromagnetic, (d) $90^\circ$-compass.  The centers of the pumping spots are shown by white dashed circles on the numerical density profiles (right column). The parameters of the numerical simulations of Eqs. (\ref{Initial GL equation},\ref{Initial Reservoir equation}) are listed in the Supp. Mat.}
     \label{summary}
      \end{figure}

{\it Equidistant vertices across a circle:} we consider a geometry of $n$ incoherently pumped equidistant polariton vertices positioned on the circumference of a circle. For equal separation distances $d=d_{ij}$  between adjacent sites the $XY$ Hamiltonian to minimise becomes ${\cal H}_{XY}=-J\sum_{i=1}^{n}\cos (\theta_{i,i+1}),$  where $J=J_{ij}$,  the summation is cyclic  and we took into account only nearest neighbour interactions. If  $J$ is positive, then all sites lock in phase ($\theta_{i,i+1}=0$). If $J$ is negative, the minimum of ${\cal H}_{XY}$ occurs for $\theta_{i,i+1}= \pm \pi$, when $n$ is even and for $\theta_{i,i+1}= \pm\pi (n\pm1)/n$ when $n$ is odd ($n>1$). We experimentally access these two regimes through incoherent injection of polaritons at the vertices of a square;  Figure 3(a,b,c) show the spin configuration, experimental results of the real-space tomography of the photoluminescence intensity at the energy of the condensate  at condensation threshold  and numerical simulation for a square with lattice constants that lead to anti-ferromagnetic, ferromagnetic and the next anti-ferromagnetic coupling respectively. Similar to the Ising polariton chain the type of coupling is clearly distinguishable by the number and symmetry of fringes between the vertices: zero or even for  anti-ferromagnetic (Fig.3(a,c)) and odd for ferromagnetic coupling (Fig.3(b)). These observations are in agreement with the $\pi$ phase difference reported in Ref.\cite{tosi13}. We can thus summarise in the case of the square lattice cell that for ferromagnetic coupling polaritons at the vertices lock with zero phase difference and for anti-ferromagnetic coupling polaritons at neighbouring vertices lock with a $\pi$ phase difference.

{\it $90^\circ$ compass model:} in the context of topological quantum computing apart from the trivial all ferromagnetic or all anti-ferromagnetic coupling configurations in a square geometry, more complex coupling configurations are of interest. Examples of such configurations are the compass models, where the coupling between the internal spin components is inherently directionally dependent. Such compass-type coupling appears in various  physical systems, where the interactions are sensitive to the spatial orientation of the involved orbitals. In polariton graphs the compass models with direction dependent coupling or spin glassy models with random couplings can be realised by changing the pumping intensity and preserving the square geometry, or alternatively, tuning the separation distances so that each vertex has one ferromagnetic and one anti-ferromagnetic coupling with its nearest neighbours. In Fig. 3(c) we have realised the $90^\circ$ compass model, where each vertex has one ferromagnetic and one anti-ferromagnetic coupling with its neighbours as it is clearly distinguishable by the number of fringes between nearest vertices. The $90^\circ$ compass, where both ferro- and anti-ferromagnetic coupling appear across the two orthogonal diagonals here, has been proposed as a model to Mott insulators with orbital degrees of freedom and frustrated magnets \cite{Nussinov2015}. Other compass-type models accessible through polariton graphs include the plaquette orbital model, where the ferromagnetic and anti-ferromagnetic coupling alternate along each direction \cite{biskup2010} and the orbital compass model on a checkerboard lattice \cite{nasu2012}. Fully random couplings in the square lattice describes the thermodynamic behaviour of several disordered systems, such as magnetic systems with random Dzyaloshinskii- Moriya interactions \cite{Rubinstein1983},  disordered Josephson junction arrays  \cite{kosterlitz1989},  disordered substrates \cite{cha}, and vortex glasses in high-$T_c$ cuprate superconductors \cite{gingras1996}. It should be possible to address these systems by considering a  polariton lattice composed of  the individual compass elements analogous to the one we realised here.

 %\begin{figure}[]
%\centering
%  \includegraphics[width=3.in]{four}      \caption{ The experimentally (left panels) and theoretically (central panels) found density profiles of the condensates pumped at the corners of a square with sides (a) $3\mu m$, (b) $4\mu m$ and (c) $5\mu m$. The right panels show the theoretically found velocity streamlines indicating the central source and four stagnation points. The  condensate wavefunctions are found by  numerical integration of the dimensionless Eq. (\ref{Initial GL equation}) for $p({\bf r})=5 \exp[-0.5r^2], b=g=1$.}
 %    \label{four}
 %     \end{figure}

\begin{figure}[ht!]
\centering
  \includegraphics[width=8.6cm]{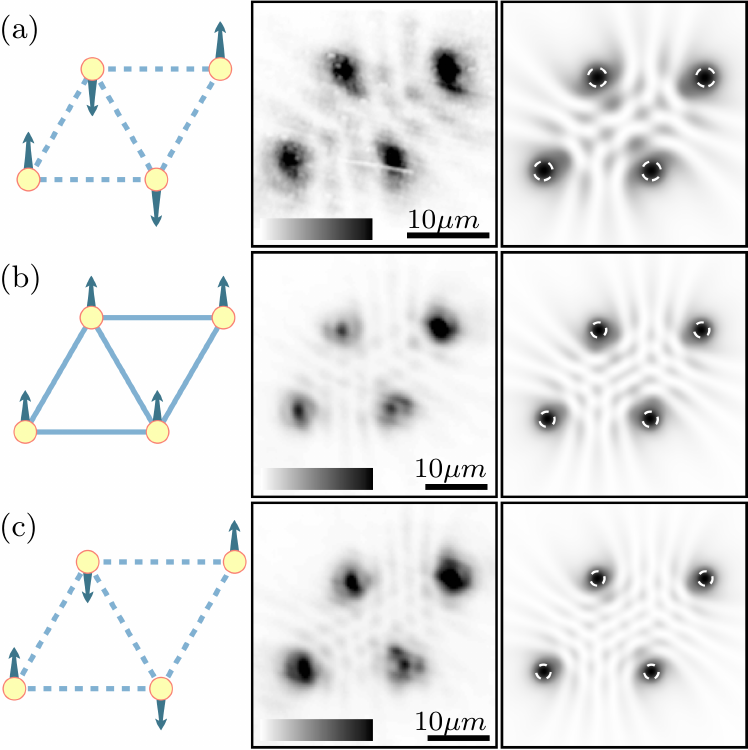}   	
\caption{ Spin configurations of the diamond-shaped polariton lattices.  The columns of images are as described in the caption to Fig.3.  The configurations shown are some elementary building blocks of triangular lattices such as  (a,c) anti-ferromagnetic and (b) ferromagnetic rhombuses. The  false-grey scale images of the middle column show the normalised photoluminescence intensity of the real-space tomography at the energy of the condensate saturated at 0.5 to increase the visibility of the low intensity fringes between the vertices.}
     \label{summary}
      \end{figure}

{\it Triangular lattice:} the $XY$ Hamiltonian has been simulated on a triangular lattice of atomic condensates discovering variety of magnetic phases and frustrated spin configurations \cite{struck11}. In the case of an anti-ferromagnetically coupled polariton triad, arranged at the vertices of an equidistant triangle, the energy flux that minimizes the $XY$ Hamiltonian corresponds to $\pm1$  winding ($2\pi/3$ phase difference between the condensates) \cite{ohadi14}. Here, we experimentally realise an equidistant triangular lattice of two  lattice cells (rhombus configuration) under incoherent injection of polaritons in the bespoke lattice configurations. Figure 4(a,b,c) show the spin configuration, experimental results of the real-space tomography of the photoluminescence intensity at the energy of the condensate  at condensation threshold and numerical simulation for a rhombus with lattice constants that lead to anti-ferromagnetic, ferromagnetic and the next anti-ferromagnetic coupling respectively. In the case of ferromagnetic coupling between nearest neighbours and neglecting opposite neighbours interaction across the long diagonal axis of the rhombus, the $XY$ Hamiltonian is minimised at ${\cal H}_{XY}\sim-5J $ when all polariton sites lock in phase, as shown in Fig.~4(b). Similarly, in the case of anti-ferromagnetic coupling between nearest neighbours the $XY$ Hamiltonian is minimised at ${\cal H}_{XY}\sim-3J$ when there is $\pm\pi$ phase difference between the outer edges of the rhombus. This configuration forces the rhombus in a frustrated state wherein opposite vertices have the same phase. This type of frustrated spin configuration is experimentally realised in Fig. 4(a,c).   The corresponding states in Figs. 4(a,b,c)  are shown in the order of the increasing distance between the sites, therefore, the anti-ferromagnetic states of Figs. 4(a) and 4(c) belong to two different bands of anti-ferromagnetic regions separated by a ferromagnetic band (the alternating anti-ferromagnetic/ferromagnetic couplings bands are shown in Fig. 2(a)). The measured density profiles show some clear differences: the local minimum at the center of the rhombus along the long diagonal in Fig. 4(a) is replaced by a local maximum in Fig. 4(c). %The appearance of a local maximum or minimum agrees with  a linear superposition of four cylindrical waves with a given $k_c$ and phase shift $\phi$, with a phase configuration as dictated by the minimisation of the $XY$ Hamiltonian, that results in a (partial) constructive or destructive interference at the centre of the rhombus.

\begin{figure}[t]
\centering
  \includegraphics[width=8.6cm]{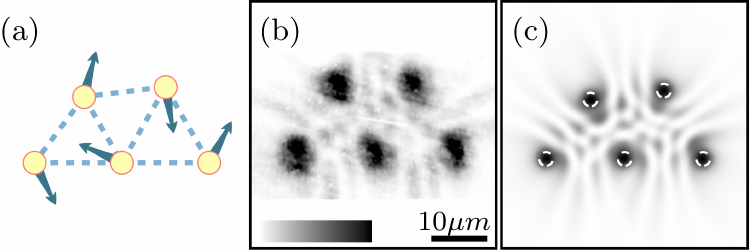}
\caption{ Spin configurations of a random polariton graph. The panels of images are as described in the caption to Fig.3. The  false-grey scale image of the middle column show the normalised photoluminescence intensity of the real-space tomography at the energy of the condensate saturated at 0.5 to increase the visibility of the low intensity fringes between the vertices.}
     \label{summary}
      \end{figure}

{\it Random polariton graph:} Beyond the minimization of the $XY$ Hamiltonian of polariton condensates on regular lattices we test our platform on a disordered polariton graph of five vertices. We took a graph initially consisting of three equidistant triangular unit cells for a lattice constant that leads to anti-ferromagnetic coupling, but with one spot breaking the symmetry. This is achieved experimentally by slightly displacing one spot on the graph. Figure 5 shows the spin configuration, experimental results of the real-space tomography of the photoluminescence intensity at the energy of the condensate  at condensation threshold and numerical simulations that correspond to this graph. For the symmetric configuration of  three equidistant triangular cells and considering only nearest neighbours interactions, the $XY$ Hamiltonian is minimised at ${\cal H}_{XY}\sim-3.86J $ with an alternating winding around each cell slightly deviating from $2\pi/3$ difference reported for a single equilateral triangle (see Supp. Mat. for details). Breaking the symmetry  leads to a different phase distribution, while maintaining the winding around each cell. The analysis of the fringes on the experimental image (with the different rows of local maxima along the two long diagonals) shows that the symmetry is explicitly broken.

 %
% \begin{figure}[]
%\centering
%  \includegraphics[width=3.in]{rhombus}      \caption{ The experimentally (left panels) and theoretically (right panel) found density profiles of the condensates pumped at the vertices of a right rombus with sides  (a) $8\mu m$ and (b) $9.5\mu m$. For ferromagnetic couplings the condensate wavefunction shows no winding around the pumping spots (a). For antiferromagnetic coupling the vortex of multiplicity $\pm 1$ formed at the center of the right triangles. The pairs $(+1,-1)$ and $(-1,+1)$ have the same number of particles and are equally likely to form. The  condensate wavefunctions are found by  numerical integration of the dimensionless Eq. (\ref{Initial GL equation}) for $p({\bf r})=5 \exp[-0.5r^2], b=g=1$.}
%     \label{rhombus}
%      \end{figure}

In conclusion, we propose polariton graphs as an analog platform for minimizing the $XY$ Hamiltonian and experimentally demonstrate its experimental implementation for simple building blocks of lattices. We demonstrated that the search for the global ground state of a polariton graph is equivalent to the minimisation of the $XY$ Hamiltonian ${\cal H}_{XY}=-\sum J_{ij} \cos\theta_{ij}$. Polariton graphs offer the scalability of optical lattices, together with the potential to study disordered systems, and to control both the sign and the strength of the coupling for each edge independently. Similar to networks of time-multiplexed OPOs, phase transitions in polariton graphs occur at the global ground state. Furthermore, polariton graphs offer the potential to quench either all or  individual vertices on variable time scales and study the complex relaxation dynamics. With the recent advances in the field of polariton condensates, such as room temperature operation \citep{organic} and condensation under electrical pumping \citep{electrical}, polariton graph based simulators offer unprecedented opportunities in addressing the $XY$ Hamiltonian and therefore  topological quantum information processing and the study of exotic  phase transitions. %Finally, we would like to note that the word ``quantum" could be attached to our proposal for a simulator to reflect the statistical nature of polariton condensates. The process of Bose--Einstein condensation is inherent to quantum statistics where a large fraction of bosons occupies the lowest quantum state, at which point macroscopic quantum phenomena become apparent. The use of the classical mean-field equations to describe the kinetics of the condensate does not negate the quantum statistic nature of its existence. At the same time, the proposed simulator has a quantum speed-up which is associated with the stimulated process of condensation. % i.e. an accelerated relaxation to the global ground quantum state.

  %%%%%%%%%%%%%%%%%%%%%%%%%%

%%%%%%%%%%%%%%%%%%%%%%%%%%%

 %%%%%%%%%% Merge with supplemental materials %%%%%%%%%%
\pagebreak
\widetext
\clearpage

\begin{center}
\textbf{\large Supplemental Material: Realizing the $XY$ Hamiltonian in polariton simulators}
\end{center}
%%%%%%%%%% Merge with supplemental materials %%%%%%%%%%
%%%%%%%%%% Prefix a "S" to all equations, figures, tables and reset the counter %%%%%%%%%%
\setcounter{equation}{0}
\setcounter{figure}{0}
\setcounter{table}{0}
\setcounter{page}{1}
\makeatletter
\renewcommand{\theequation}{S\arabic{equation}}
\renewcommand{\thefigure}{S\arabic{figure}}
\renewcommand{\bibnumfmt}[1]{[S#1]}
\renewcommand{\citenumfont}[1]{S#1}
%%%%%%%%%% Prefix a "S" to all equations, figures, tables and reset the counter %%%%%%%%%%
\section*{Microcavity Sample}
The semiconductor microcavity structure studied here is a planar, strain compensated 2$\lambda$ GaAs microcavity with embedded InGaAs quantum wells (QWs). Strain compensation was achieved by AlAs$_{0.98}$P$_{0.02}$/GaAs DBR layers instead of the thin AlP inserts in the AlAs layers used in Ref.\,[\onlinecite{suppression_2012}] as their effective composition could be better controlled. The bottom DBR consists of 26 pairs of GaAs and AlAs$_{0.98}$P$_{0.02}$ while the top has 23 of these pairs, resulting in very high reflectance (\textgreater 99.9$\%$) in the stop-band region of the spectrum. The average density of hatches along the [$110$] direction was estimated from transmission imaging to be about 6/mm, while no hatches along the [$1\bar{1}0$] direction were observed. Three pairs of 6\,nm In$_{0.08}$Ga$_{0.92}$As QWs are embedded in the GaAs cavity at the anti-nodes of the field as well as two additional QWs at the first and last node to serve as carrier collection wells.  The large number of QWs was chosen to increase the Rabi splitting and keep the exciton density per QW below the Mott density \cite{saturation} also for sufficiently high polariton densities to achieve polariton condensation under non-resonant excitation. The strong coupling between the exciton resonance and the cavity mode is observed with a vacuum Rabi-splitting of $2\hbar\Omega\sim8$\,meV. A wedge in the cavity thickness allows access to a wide range of exciton-cavity detuning.  All measurements reported here are taken at  $\Delta\approx -5.5$\,meV. The measured Q-factor is $\sim 12000$, while the calculated bare cavity Q-factor, neglecting in-plane disorder and residual absorption, is $\sim25000$. As the emission energy of the InGaAs QWs is lower than the absorption of the GaAs substrate we can study the photoluminescence of the sample both in reflection and transmission geometry. The transmission geometry, which is not available for GaAs QWs, allows to filter the surface reflection of the excitation, and has been widely utilized to probe the features of polariton fluids \cite{all-optical_2011,nardin_hydrodynamic_2011} under resonant excitation of polaritons. Using real and reciprocal space spectroscopic imaging under non-resonant optical excitation, polariton condensation, and a second threshold marking the onset of photon lasing, i.e. the transition from the strong to the weak-coupling regime has been studied in this microcavity \cite{InGaAs}.

\section*{Experimental setup}
In the experiments described here the sample was held in a cold finger cryostat at a temperature of~$T\approx 6$\,K. Continuous wave excitation is provided by a Ti:Sapphire laser. We use non-resonant excitation from the epi side, and detect the emission from the substrate side, so that the excitation is filtered by the absorption of the GaAs substrate. The optical excitation, for all the measurements reported in this work, is at the first reflectivity minimum above the cavity stop band. The spatial profile of the excitation beam is modulated to a graph with Gaussian profiles at each vertex of approximatelly equal in diameter spots using a reflective spatial light modulator (SLM). We use a high numerical aperture microscope objective (NA = 0.65) to focus the spatially modulated beam to $\sim $1-2$\, \mathrm{\mu m}$ in diameter at full width at half maximum (FWHM) excitation spots. The photoluminescence from the sample is collected in transmission geometry with $\pm 25^{\circ}$ collection angle, by a 0.42 NA microscope objective.  Fourier (dispersion) imaging is performed by projecting the Fourier-space at the slit of a $300$\,mm spectrophotometer coupled a cooled charge coupled (CCD) device and using a $1200$ grooves/mm with 50\, $\mu$eV energy-resolution. The real-space tomography images are acquired with sub-micron optical resolution using a CCD camera imaging configuration through a tunable Fabry-Perot etalon with $\sim$ 20\, $\mu$eV FWHM bandwidth.   

\section*{Wavevector Tomography}

The condensate wavevector, $k_c$, of the polariton Ising chains is measured using two dimensional Fourier-space tomography utilising a tunable Fabry-Perot etalon with $\sim$ 20\, $\mu$eV FWHM bandwidth and a CCD camera imaging configuration. Figure 1(a,b) in Supp. Mat. shows the false-colour normalised photoluminescence intensity of the two dimensional cross-section of the Fourier-space  at the energy of the condensate from the Ising chain configuration of Fig.2(b) and Fig.2(c) respectively at condensation threshold. The outer ring in both images corresponds to $k_c$, whereas the inner fringes correspond to self-diffraction from the Ising chain. For the anti-ferromagnetic Ising chain of Fig.2(a), $k_c \approx$1.79\,$\mu m^{-1}$ and for the ferromagnetic Ising chain of Fig.2(c), $k_c \approx$1.67\,$\mu m^{-1}$.   

\begin{figure}[h]
\centering
  \includegraphics[width=16.cm]{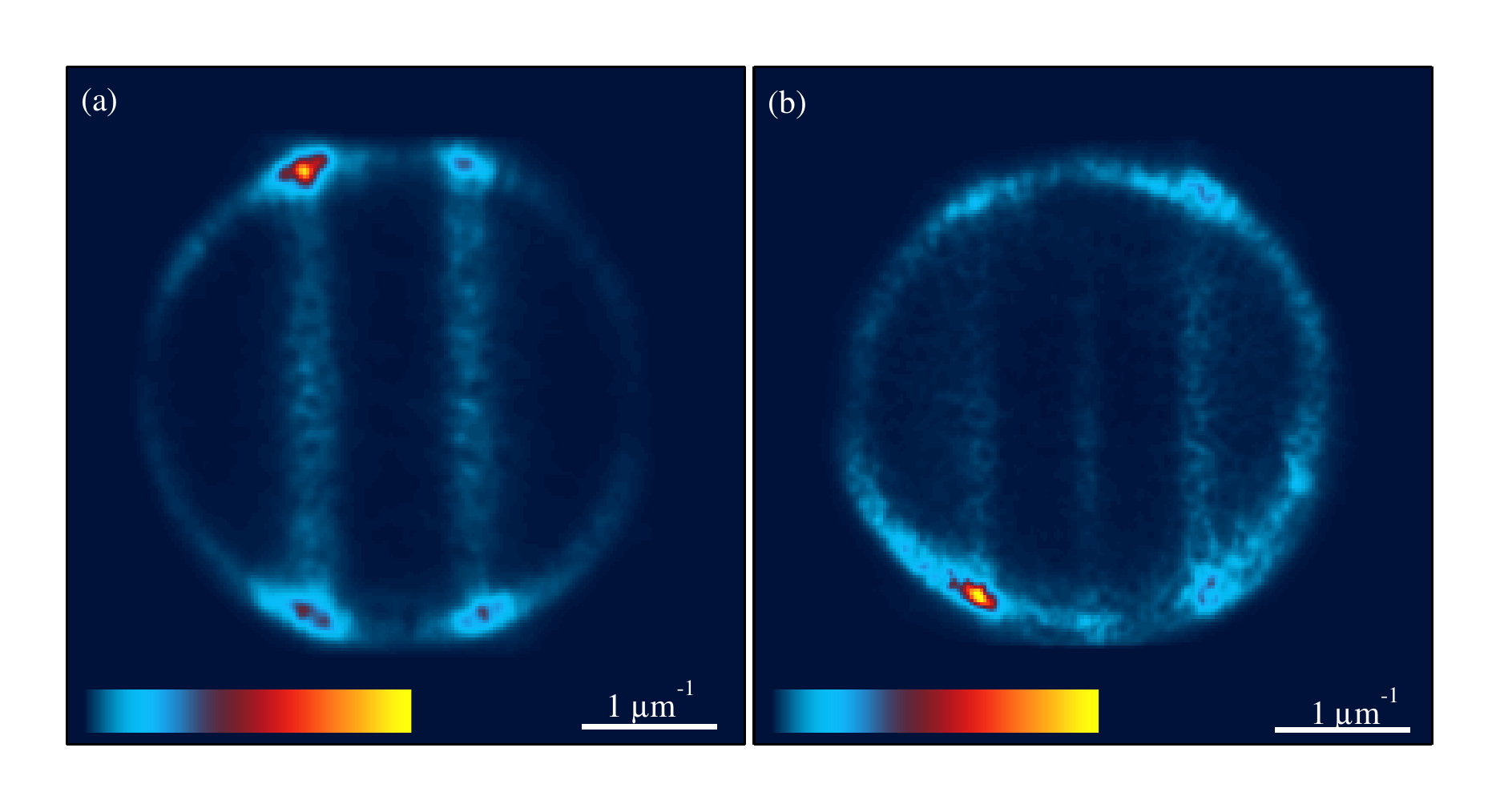}
    	
    \vspace{0cm}
       \caption{(a) False-colour normalised photoluminescence intensity of the Fourier-space tomography at the energy of the condensate from the Ising chain configuration of Fig.2(b). (b) same as (a) but for the the Ising chain configuration of Fig.2(c). }
     \label{KK_suppmat}
      \end{figure}

\section*{Finding the expression for the coupling coefficients}
The expression for the coupling coefficients $J_{ij}$ given by Eq. (8) can be estimated  based on the width of the Hankel transformation  of the wavefunction of an individual condensate given by 
\begin{equation}
\widehat{\Psi}(k)=2\pi\int_0^\infty\sqrt{\rho(r)} \exp[i k_c r]J_0(kr)r\,dr.
\label{A1}
\end{equation}
The density of the Hankel transformation, $|\widehat{\Psi}(k)|^2$,  peaks at $k=k_c$ with the width, characterized by $\epsilon$,  inversely proportional to the width of the condensate density $\rho(r)$, which is set by the width of the pumping profile $p(r)$.  We, therefore, approximate $|\widehat{\Psi}(k)|^2$ by
\begin{equation}
|\widehat{\Psi}(k)|^2\approx |\widehat{\Psi}(k_c)|^2 \frac{{\rm rect}(\frac{k-k_c}{\epsilon})}{\epsilon}.
\label{A2}
\end{equation}
We integrate Eq. (8) using Eq. (\ref{A2}) to get 
\begin{eqnarray}
J_{ij}&=&\frac{1}{\pi}|\widehat{\Psi}(k_c)|^2\biggr[\biggl(\frac{k_c}{d_{ij}\epsilon}-\frac{1}{2d_{ij}}\biggr) J_1\biggl( \frac{d_{ij}\epsilon}{2}- k_c d_{ij}\biggr) \nonumber \\
&&+ \biggl(\frac{k_c}{d_{ij}\epsilon}+\frac{1}{2d_{ij}}\biggr) J_1 \biggl(\frac{d_{ij}\epsilon}{2}+ k_c d_{ij}) \biggr)\biggl].
\label{jj}
  \end{eqnarray}
  In the limit of  $\epsilon\rightarrow 0$ we recover our $\delta-$function approximation $J_{ij}= k_c \vert \widehat\Psi(k_c) \vert^2 J_0(k_c d_{ij})/\pi$. The finite width of the Hankel transformation of the condensate wavefunction, as seen from Eq. \ref{jj}, induces a phase shift, so the criterion for the phase switching can be approximated by the sign switching of $\cos(k_c d_{ij}+\phi)$, where $\phi$ is the system parameter dependent term.
  
  \section*{Minimization of the XY Hamiltonian for sample configurations}
  
  We find the global minimum of the XY Hamiltonian directly for the sample configurations considered in our paper. For the lattice sites arranged in a square the phases relative to one fixed phase that we set equal to zero, $\theta_0=0$,  minimize the XY Hamiltonian
  \begin{equation}
  {\cal H}_\square=-J(\cos \theta_{10} + \cos\theta_{12}+\cos\theta_{23}+ \cos\theta_{30}) - J\delta (\cos \theta_{20}+\cos\theta_{13}),
  \end{equation}
  where we denoted $\delta$ to be the ratio of the coupling of the diagonal cites to the coupling of the neighboring sites.  The coupling strength decays with the distance between sites, therefore, $|\delta|<1$. If all couplings are ferromagnetic, $J, \delta>0$, the minimum of ${\cal H}_\square$ is for $\theta_{i0}=0$. If $J<0$, there is a $\pi$ phase difference between the neighboring sites $\theta_{10} =\pi, \theta_{20}=0, \theta_{30}=\pi$ even for nonzero $\delta$ (as long as $|\delta|<1$ is satisfied).
  
  For a rhombus, consisting of two equilaterial triangles,  the XY Hamiltonian becomes
  \begin{equation}
  {\cal H}_{rh}=-J(\cos \theta_{10} + \cos\theta_{20}+\cos\theta_{30}+ \cos\theta_{12}+\cos\theta_{23}) - J\delta\cos\theta_{13},
  \label{hrhombus}
  \end{equation}
  where we associated $\theta_0=0$ with one of the sites along the shorter diagonal. $\delta$ in this case represents the ratio of the coupling along the long diagonal to that between the neighbours. While for an equilaterial triangle the XY Hamiltonian  ${\cal H}_\triangle=-J (\cos\theta_{10}+\cos\theta_{20}+\cos\theta_{12})$, $J<0$  is minimized by $\theta_{i0}=\pm 2\pi/3$, the XY Hamiltonian (\ref{hrhombus}) is minimized by $\theta_{10}=\theta_{30}=\pi$ , $\theta_{20}=0$ as in the case of the square. 
  
  For three equilaterial triangles non-trivial winding around sites is again possible, since the XY Hamiltonian 
  \begin{equation}
  {\cal H}_5=-J(\cos \theta_{10} + \cos\theta_{20}+\cos\theta_{30}+ \cos\theta_{40}+\cos\theta_{12}+\cos\theta_{23}+\cos\theta_{34})
  \label{h5}
  \end{equation}
  for $J<0$ 
    reaches its minimum at $\theta_{10}=-\theta_{40}=\pm 0.73\pi, \theta_{20}=-\theta_{30}=\mp0.54\pi$, therefore, creating an alternating winding around each of the equilaterial triangles. Here we associated $\theta_0=0$ with the site that is connected to all other sites and neglected the interactions along two long diagonals.  If the distances are close to the switching points between ferro- and antiferro- couplings the small deviation in the position of the sites may lead to an even more complex configurations as is illustrated on Fig. 5 of the main text. 
  
   \section*{Parameters of the numerical simulations}
   
  In our numerical simulation we used a Gaussian pumping profile that produces the same width of the condensate as in experiment  (FWHM  $2.6 \mu m$) and choose the pumping intensity to obtain the correct outflow wavenumber for a single condensate. The common integration parameters used for all numerical simulations are, therefore,    
 $g = 0.1, b = 1, \gamma = 0.3, \ \eta = 0.4, \ p = 9.5 \exp(-0.4 r^2).$  The numerical simulations were performed for various geometries and distances as the main text shows. For Fig. 3 we varied the distances between the nearest neighbors, so that $k_c d = 12.1, 15.3, 18.4, 16.8$ for Figs. 3a,b,c,d respectively. For Fig. 4 we used $k_c d=19, 22.3, 24.9$ for Figs. 4a,b,c respectively. For Fig.5 we used $k_c d = 18.2\pm0.5$.

\end{document}